# Bound Eigenstate dynamics under a sudden shift of the well's wall

Er'el Granot and Avi Marchewka

*Department of electrical and Electronics Engineering,*
*Ariel University Center of Samaria, Ariel, Israel*

**Abstract**:

We investigate the dynamics of the eigenstate of an infinite well under an abrupt shift of the well's wall. It is shown that when the shift is small compared to the initial well's dimensions, the short time behavior changes from the well known $t^{3/2}$ behavior to $t^{1/2}$. It is also shown that the complete dynamical picture converges to a universal function, which has fractal structure with dimensionality $D = 1.25$.

## Introduction:

When a smooth wavefunction evolve in time the change in the probability density and in the survival probability is proportional to $t^2$. However, in numerous scenarios either when the initial wavefunction or its derivatives are discontinuous this is not the case. It is well known that the temporal behavior of the probability density of an initially non analytical function has a fractional power law at short times. A particular scenario is the Moshinsky's Shutter[1] or a collapse of the wavefunction after spatial measurement[2]. Moreover, it was shown[3] that the initial wavefunction does not have to be discontinuous to present a discontinuous behavior, even a sharp but smooth wavefunction demonstrate at certain conditions a discontinuous behavior.

In particular, the survival probability of an initial state, which has a discontinuity, has a $t^{1/2}$ dependence at short times[3]. Similarly, if the initial wavefunction is continuous but has a discontinuity in its derivative, has a $t^{3/2}$ dependence at short times[3].

Therefore, it is not surprising that when the eigenstate of an infinite well is released by widening the well a little, the initial survival probability has a $t^{3/2}$ dependence on time, since the eigenstate has a discontinuous derivative (while the wavefunction itself is continuous). However, it will be shown that this is only a partial picture. This temporal dependence can remain only at free space. When the wavefunction eventually reaches the new boundary of the well, the reflections will reduce the rate in which the survival probability decays and the survival probability will decay likes $t^{1/2}$ as if the initial wavefunction was discontinuous and not its derivative (for others initially non-analytical decay processes, see Refs.4-8).

Moreover, initially non-analytical wavefunctions demonstrate fractal dynamics. This behavior was demonstrated in many works[9-14]. The fractal dimension depends on the spectral distribution of the initial state. In this work we calculate the escape probability (the probability to escape form the initial state), and show that when the wall shift is miniscule the fractal nature does not vanish, on the contrary, although its amplitude decreases its shape is universal, and is independent of the shift.

**The model:**

Accordingly, we investigate the following Schrödinger equation

$$i\frac{\partial}{\partial t}\psi = -\frac{\partial^2}{\partial x^2}\psi + V(x,t) \qquad (1)$$

where we adopt the units: $2m = 1$, $\hbar = 1$ and

$$V(x,t) = \begin{cases} 0 & 0 < x < l(t) \\ \infty & \text{else} \end{cases} \qquad (2)$$

is the well's potential when

$$l(t) = \begin{cases} 1 & t \le 0 \\ L = 1 + \Delta & t > 0 \end{cases}$$

is the varying width of the barrier.

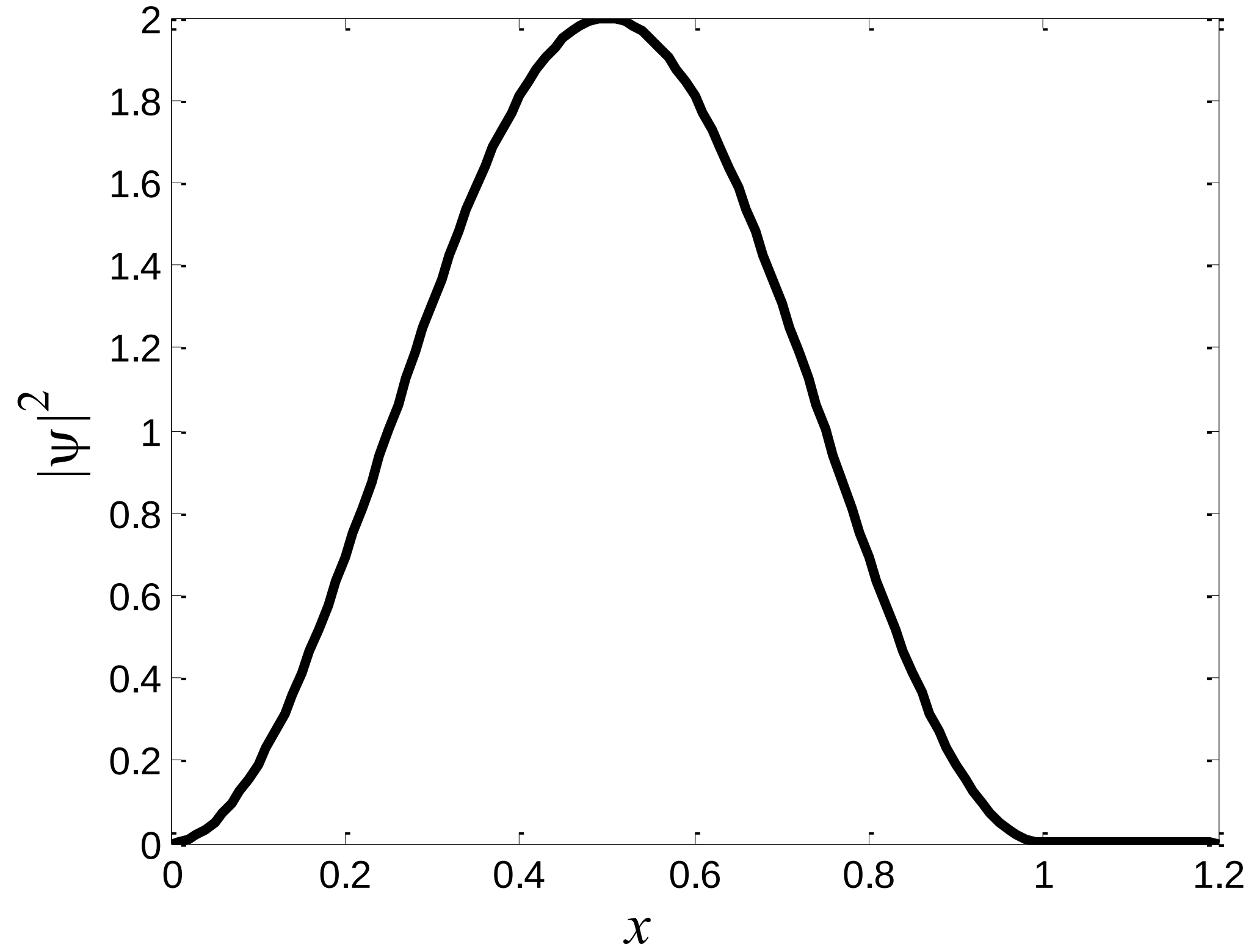

Figure 1: The plot of the eigenstate at $t > 0$ (after the spatial shift of the well's right wall). In this case $\Delta = 0.2$.

The initial eigenstate is therefore (Fig.1):

$$\psi(x, t \le 0) = \sqrt{2}\sin(\pi x) \tag{3}$$

while for $t > 0$ it can be written as a series:

$$\psi(x, t > 0) = \sqrt{\frac{2}{L}} \sum_{n=1}^{\infty} a_n \sin\left(n\frac{\pi x}{L}\right) \exp\left[-i\left(\frac{\pi n}{L}\right)^2 t\right]$$

where coefficients are

$$a_n = \frac{2}{\sqrt{L}}\int_0^1 dx \sin(\pi x)\sin\left(n\frac{\pi x}{L}\right) = \frac{2}{\pi\sqrt{L}}\sin\left(\frac{\pi n}{L}\right)\frac{1}{1-(n/L)^2}$$

therefore

$$\psi(x,t>0) = \frac{\sqrt{8}}{\pi L}\sum_{n=1}^{\infty}\sin\left(\frac{\pi n}{L}\right)\frac{1}{1-(n/L)^2}\sin\left(n\frac{\pi x}{L}\right)\exp\left[-i\left(\frac{\pi n}{L}\right)^2 t\right] \qquad (4)$$

This problem was addressed in the literature using various techniques [15-19]. It was shown that the system exhibit periodic behavior, which is independent of the initial system dimensions. The period was found to be $T = 2\pi\hbar / E_1$, where $E_1$ is the ground state energy of the expanded system. In our units $T = 2L^2/\pi$.
Moreover, at any rational fraction of $T$ the initial wavefunction is replicated creating an intricate structure. At non-rational fraction of $T$ the wavefunction is therefore not analytic at any point. The wavefunction is then fractal.

In Fig.2 we present the dynamics of $|\psi(x,t)|^2$ for a single period $T = 2L^2/\pi$ ( approx. 0.9167 in this case) . In this figure the gray levels correspond to the value of $|\psi(x,t)|^2$, where white stands for 0 and back stands for 2. The periodicity and the intricate structures are clearly seen. In this manuscript, however, we will focus only on the short time and the $\Delta \to 0$ regimes.

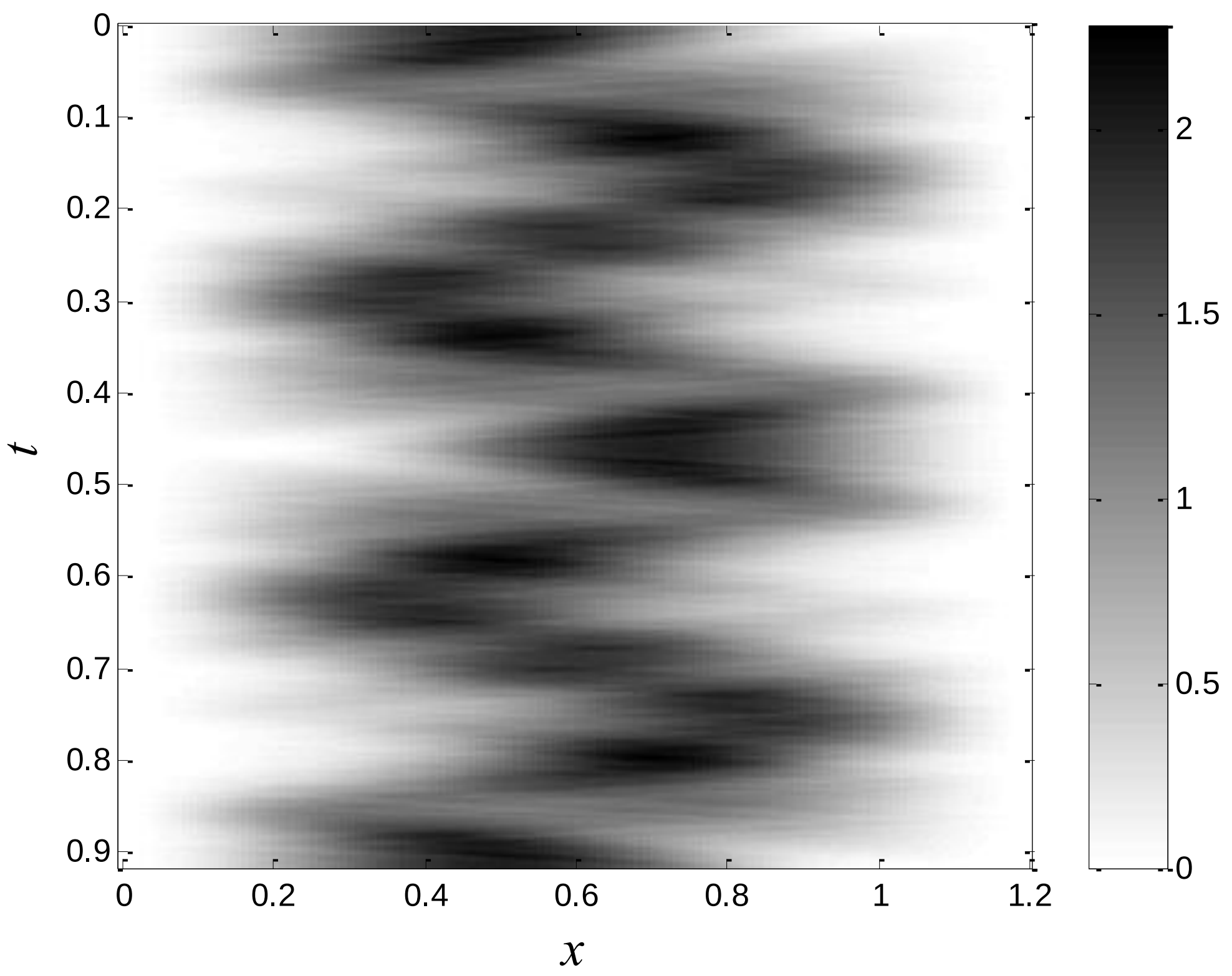

Figure 2: The spatial and temporal dynamics of probability density for expansion $\Delta = 0.2$, where black corresponds to the maximum (2), and white to the minimum value (0).

**The escaping probability and its approximations**

The survival amplitude reads:

$$A(t) = \langle \psi(x,t) \,|\, \psi(x,0) \rangle = \frac{4}{\pi^2 L} \sum_{n=1}^{\infty} \sin^2\left(\frac{\pi n}{L}\right) \frac{1}{\left[1-(n/L)^2\right]^2} \exp\left[-i\left(\frac{\pi n}{L}\right)^2 t\right] \quad (5)$$

and the corresponding survival probability is therefore:

$$P(t) = |A(t)|^2 \quad (6)$$

We also define the escape probability, which can be evaluated

$$P_{escape}(t) = 1 - P(t) = 1 - \left| \frac{4}{\pi^2 L} \sum_{n=1}^{\infty} \sin^2\left(\frac{\pi n}{L}\right) \frac{1}{\left[1-(n/L)^2\right]^2} \exp\left[-i\left(\frac{\pi n}{L}\right)^2 t\right] \right|^2 =$$

$$1 - \left| 1 + \frac{4}{\pi^2 L} \sum_{n=2}^{\infty} \sin^2\left(\frac{\pi n}{L}\right) \frac{1}{\left[1-(n/L)^2\right]^2} \left( \exp\left[-i\left(\frac{\pi n}{L}\right)^2 t\right] - 1 \right) \right|^2 \tag{7}$$

At short times, i.e., $t << 1$, the escape probability can be approximated

$$P_{escape}(t) \cong -2\Re\left\{ \frac{4}{\pi^2 L} \sum_{n=2}^{\infty} \sin^2\left(\frac{\pi n}{L}\right) \frac{1}{\left[1-(n/L)^2\right]^2} \left( \exp\left[-i\left(\frac{\pi n}{L}\right)^2 t\right] - 1 \right) \right\}$$

The one in the denominator can be neglected since in the short time regime the main contribution comes from the large (and many) $n$ terms and therefore

$$P_{escape}(t) \cong -\frac{8L^3}{\pi^2} \sum_{n=2}^{\infty} \sin^2\left(\frac{\pi n}{L}\right) \frac{1}{n^4} \left( \cos\left[\left(\frac{\pi n}{L}\right)^2 t\right] - 1 \right) \tag{8}$$

When the spatial shift in the wall's position is small with respect to its initial width $L = 1 + \Delta$, i.e., $\Delta << 1$, further approximations can be taken

$$P_{escape}(t) \cong \frac{16}{\pi^2} \sum_{n=2}^{\infty} \sin^2(\pi n \Delta) \frac{\sin^2\left[(\pi n)^2 t / 2\right]}{n^4} \tag{9}$$

At short time, the summation is equivalent to integration over the continuous variable $y \equiv \pi n \sqrt{t}$ :

$$P_{escape}(t) \cong 16\pi t^{(3/2)} \int_0^{\infty} dy \frac{\mathbf{sin}^2\left(y \frac{\Delta}{\sqrt{t}}\right)}{y^4} \mathbf{sin}^2\left(\frac{y^2}{2}\right) \tag{10}$$

In general, this integral can be evaluated by complex Hypergeometric functions, however, for two temporal regimes, it can be simplified considerably:

We find a transition time, which occurs approximately at $\sim \Delta^2$.

When $t << \Delta^2$ then

$$P_{escape}\left(t << \Delta^2\right) \cong 8\pi t^{(3/2)} \int_0^\infty dy \frac{\mathbf{sin}^2\left(y^2/2\right)}{y^4} = \frac{8}{3}\frac{\pi^{3/2}}{\sqrt{2}} t^{3/2} \tag{11}$$

However, beyond the transition time, i.e., for $\Delta^2 << t << 1$

$$P_{escape}(t) \cong 16\Delta^2 \pi t^{1/2} \int_0^\infty dy \frac{\mathbf{sin}^2\left(y^2/2\right)}{y^2} = 8\Delta^2 \frac{\pi^{3/2}}{\sqrt{2}} t^{1/2} \tag{12}$$

The transition time ($\sim \Delta^2$) is the time it takes the wavefunction to reach the new position of the well's wall. Prior to this time, the dynamics resembles a free space scenario, and therefore, the escape probability goes like $t^{3/2}$ [3]; however, at longer times, when the transition time $\sim \Delta^2$ is passed the increase in the escape probability reduces to $t^{1/2}$. In Fig.3 the temporal transition is shown for $\Delta = 0.003$, which corresponds to a transition time of $\sim 10^{-5}$.

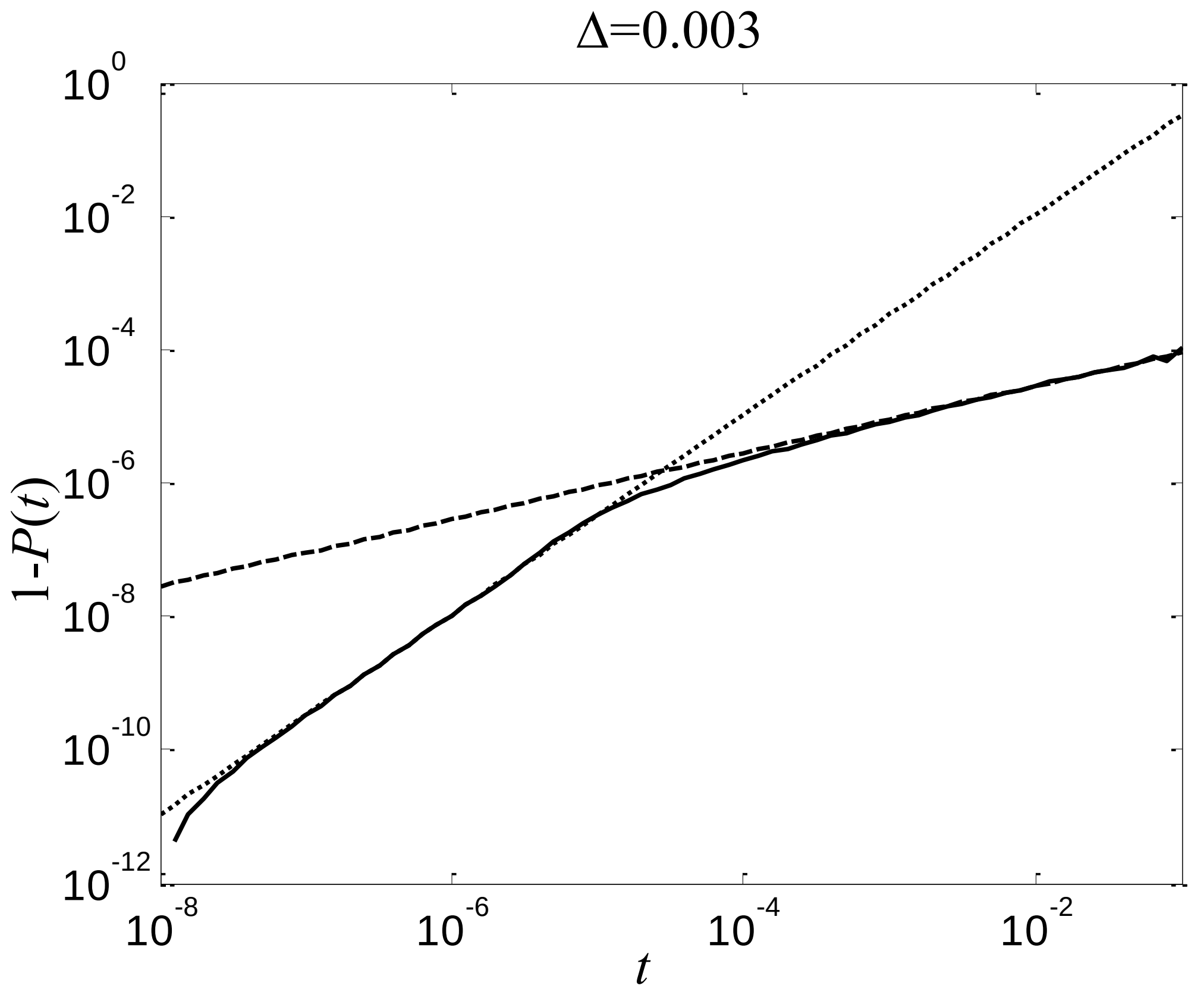

Figure 3: the plot of the escape probability as a function of time (solid line). The $t << \Delta^2$ regime approximation (Eq.11) is presented by the dotted line, while the $\Delta^2 << t << 1$ regime approximation (Eq.12) is presented by the dashed line.

Clearly, the two regimes are equal at $t = 3\Delta^2$.

When $\Delta \to 0$ the $t^{3/2}$ dependence disappear, as was explained above, and the $t^{1/2}$ dependence rules for short times; however, in this regime, it can be shown that this behavior converges to temporal behavior, which is generic for any time period.

**Fractal behavior**

In the regime $\Delta \to 0$ then $P_{escape}(t) \to 0$ as well, which means that the same approximation can be taken, yielding

$$P_{escape}(t) \cong 8\Delta^2 \sum_{n=2}^{\infty} \frac{n^2}{\left[1-n^2\right]^2} \left(1 - \cos\left[2\pi n^2 \frac{t}{T}\right]\right) \tag{13}$$

which can be rewritten as

$$P_{escape}(t) \to 8\Delta^2 F\left(\frac{t}{T}\right) \tag{14}$$

where $T \equiv 2L^2/\pi$ is the retrieval time period

and

$$F(\xi) \equiv \sum_{n=2}^{\infty} \frac{n^2}{\left(1-n^2\right)^2}\left[1-\cos\left(2\pi n^2 \xi\right)\right] \tag{15}$$

is a *universal function,* which has a fractal structure. This function is universal in the sense that it is independent of either $T$ or $\Delta$ (and of course, independent of time). For every $t/T = q/p^2$ for integers $q$ and $p$, there is a valley in the function $F(t/T)$, see Fig. 4.

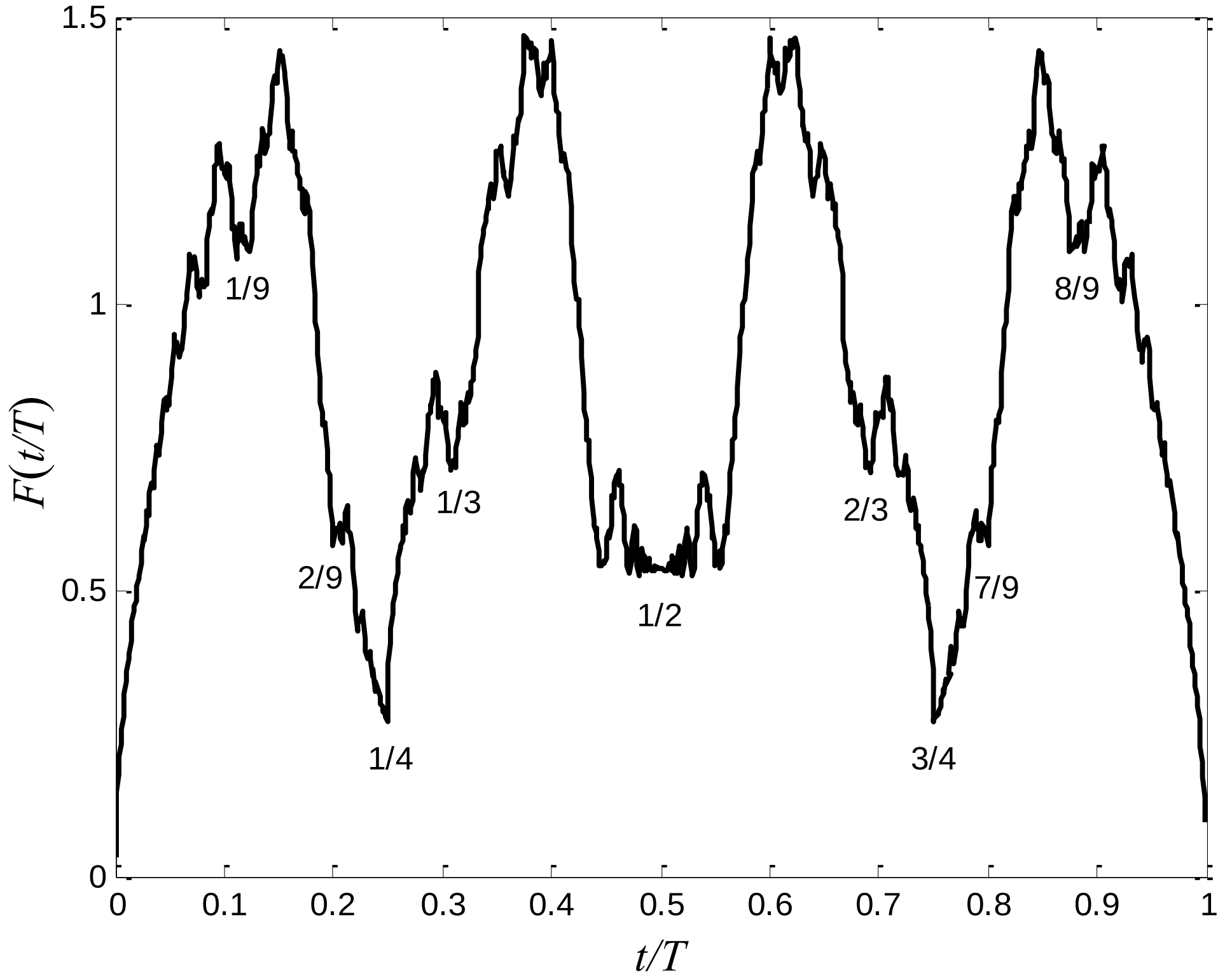

Figure 4. The temporal plot of the universal function $F(t/T)$. The main valleys (dips) in the function are signed as rational fraction $q/p^2$.

The self similarity and fractality nature of the function $F(x)$ is illustrated in Fig.5, where it is plotted at three different scales.

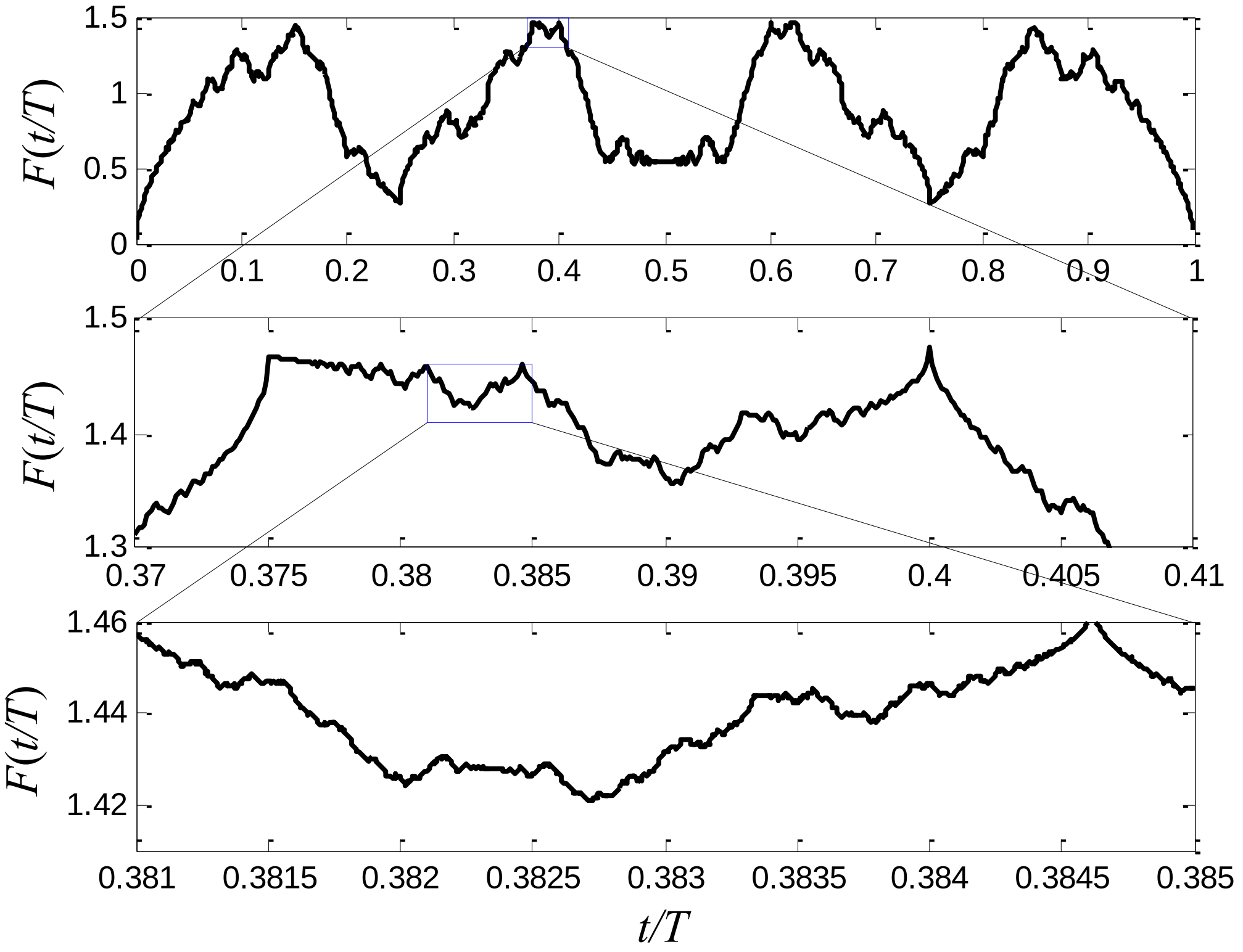

Figure:5: The fractional structure of $F(t/T)$ is illustrated by showing the similar resemblance (self similarity) of the function regardless of the scaling.

The length dimension of the curve can be evaluated as a function of the measurement unit, i.e., $\varepsilon$.

$$l(\varepsilon)=\sum_{m=1}^{1/\varepsilon}\sqrt{\varepsilon^2+\left[F(m\varepsilon+\varepsilon)-F(m\varepsilon-\varepsilon)\right]^2/4} \qquad (16)$$

The $\varepsilon^2$ term is negligible and hence

$$l(\varepsilon)\cong\frac{1}{2}\sum_{m=1}^{1/\varepsilon}\left|F(m\varepsilon+\varepsilon)-F(m\varepsilon-\varepsilon)\right|=\sum_{m=1}^{1/\varepsilon}\left|\sum_{n=2}^{\infty}\frac{n^2}{\left(1-n^2\right)^2}\left[\sin\left(2\pi n^2(m\varepsilon)\right)\sin\left(2\pi n^2(\varepsilon)\right)\right]\right|$$

Since the main contribution comes from the $n>>1$ terms then

$$l(\varepsilon)\sim\sum_{m=1}^{1/\varepsilon}\left|\sum_{n=2}^{\infty}\frac{1}{n^2}\left[\sin\left(2\pi n^2(m\varepsilon)\right)\sin\left(2\pi n^2(\varepsilon)\right)\right]\right| \qquad (17)$$

which can be rewritten as

$$I(\varepsilon) \sim \sum_{m=1}^{1/\varepsilon} 2\pi\varepsilon \left| \sum_{n=2}^{\infty} \sin\left(2\pi n^2 (m\varepsilon)\right) \operatorname{sinc}\left(2\pi n^2 (\varepsilon)\right) \right| \tag{18}$$

where the $\operatorname{sinc} x \equiv \sin x / x$ is approximately a cut-off function, which can be approximated by

$$I(\varepsilon) \sim \sum_{m=1}^{1/\varepsilon} 2\pi\varepsilon \left| \sum_{n=2}^{\sqrt{1/2\varepsilon}} \sin\left(2\pi n^2 (m\varepsilon)\right) \right| \sim 2\pi (2\varepsilon)^{-1/4}, \tag{19}$$

the terms $\mathbf{sin}\left(2\pi n^2 (m\varepsilon)\right)$ behaves like a random variable $x_{n,m}$, which can get any value between -1 and 1. Therefore, the summation $\xi_m = \sum_{n=2}^{\sqrt{1/2\varepsilon}} x_{n,m}$ is distributed normally with a mean 0 and standard deviation $\sim \varepsilon^{-1/4}$. In Fig. 6 the comparison between the distribution histogram of $\xi_m$ and the normal distribution is shown. The sum $\sum_{m=1}^{1/\varepsilon} \varepsilon\{\cdots\}$ is basically an average operation. Hence

$$I(\varepsilon) \sim \langle |\xi_m| \rangle \sim \sqrt{\langle \xi_m^2 \rangle} = \sigma \sim \varepsilon^{-1/4} \tag{20}$$

,where $\sigma$ stands for the standard deviation of the distribution. It should be emphasized that clearly there are small deviations from normal distribution, but it does not really matter so long as the standard deviation goes like $\sim \varepsilon^{-1/4}$ the conclusion is the same.

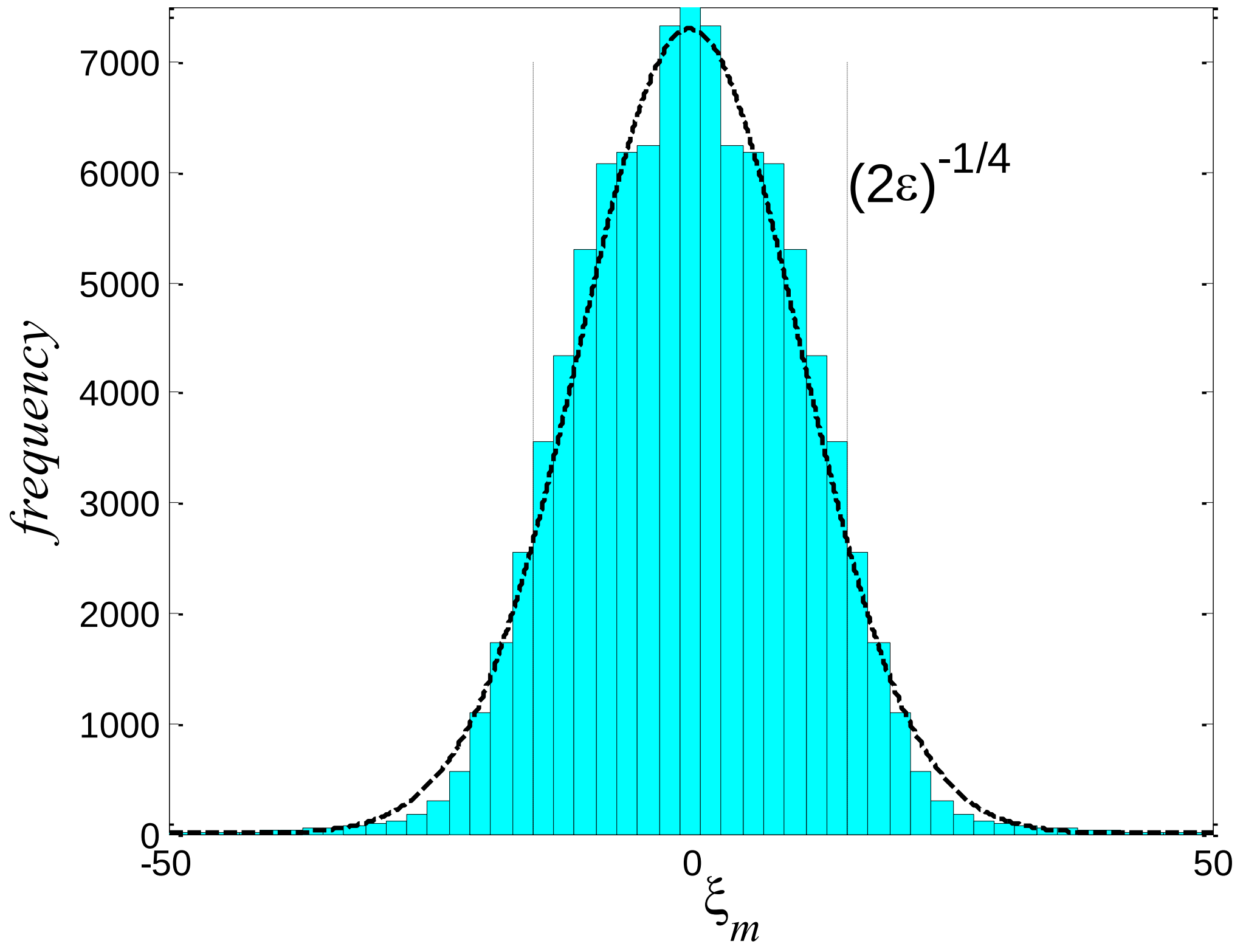

Figure 6 (color online): A histogram of the variable $\xi_m$ and its comparison to normal distribution (dashed line) for $\varepsilon = 10^{-5}$.

Therefore, the relation between the curve length and the measurement unit goes like $l(\varepsilon) \propto \varepsilon^{-1/4}$, which corresponds to a curve fractal dimension of $D = 1.25$.

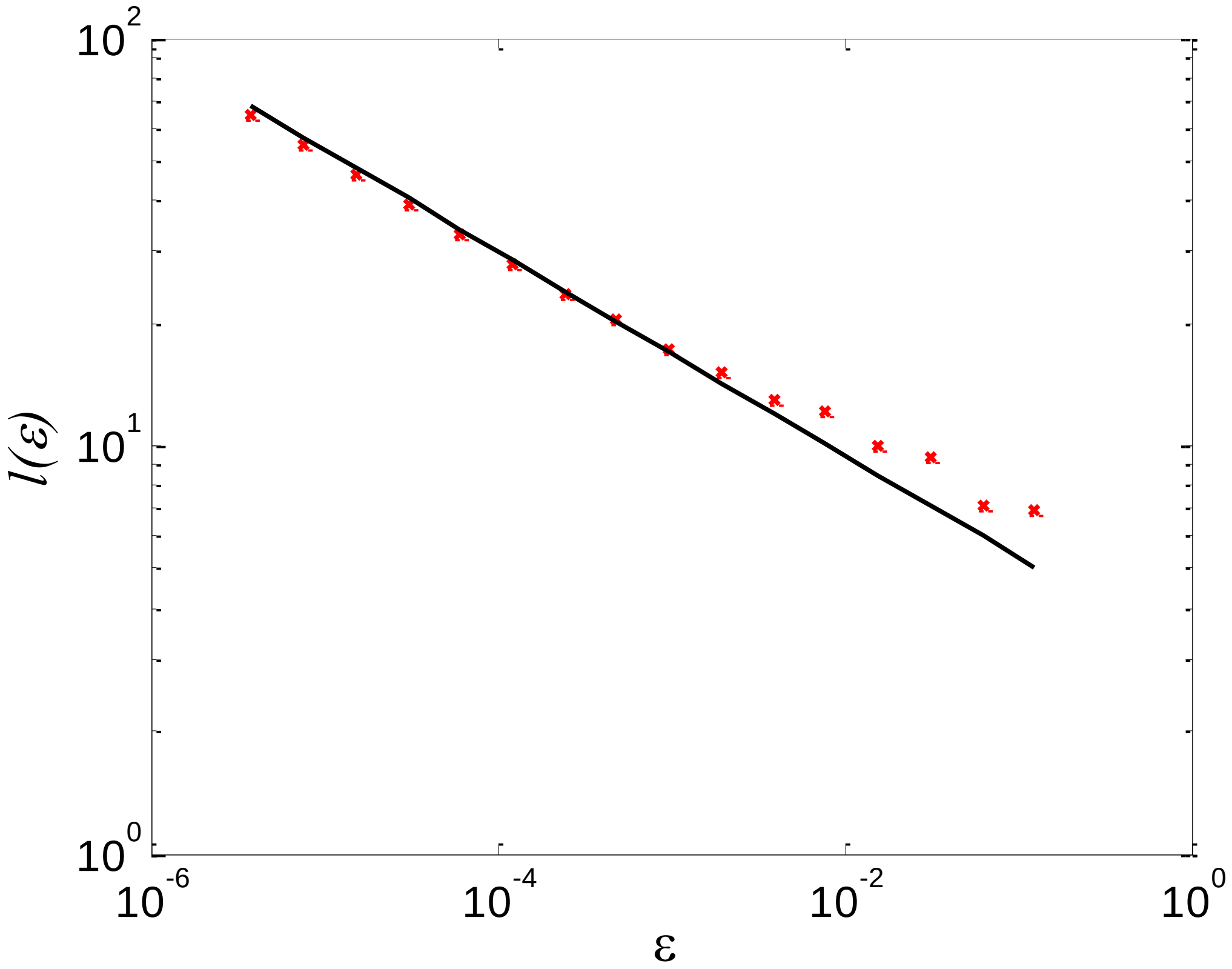

Figure 7 (color online): Numerical calculation of the length of the curve of the function $F(\xi)$ as a function of the scaling measurement $\varepsilon$ (crosses). The solid line stands for $l(\varepsilon) \propto \varepsilon^{-1/4}$ .

In Fig.7 we present the numerical evaluation of the fractal dimension of $F(x)$. As can be seen from the figure, the fractal dimension of the curve is indeed $D = 1.25$

To summarize, we have investigated the short time dynamics of the eigenstate of an infinite potential well, whose boundaries are displaced by a small distance $\Delta << 1$. It was shown that the survival probability has a transition time $t_c \cong \Delta^2$. While $t << t_c$ the escaping probability goes like $t^{3/2}$ as should be expected form the literature, however, while $t >> t_c$ the escaping probability goes like $t^{1/2}$ as if the initial wavefunction was discontinuous. Moreover, it was shown that as $\Delta \to 0$ the temporal dynamics of the escaping probability converges to a universal periodic function, which has a fractal dimension $D = 1.25$.